\documentclass{article}

\usepackage{enumerate}
\usepackage{amsfonts}
\usepackage{amsmath}
\usepackage{textcomp}
\usepackage{amsthm, tikz}
\usepackage{amssymb}
\usepackage{color}
\usepackage{graphicx}
\usepackage[matrix,arrow,curve]{xy}
\usepackage{array}
\usepackage{mathtools}
\usepackage{comment}

\usepackage{tikz}
\usetikzlibrary{arrows}
\usepgflibrary{arrows}

\newcommand{\mathpic}[2]{\ensuremath{\vcenter{\hbox{\begin{tikzpicture}[scale=#1,>=latex'] #2 \end{tikzpicture}}}}}

\def\be{\begin{eqnarray}}
\def\ee{\end{eqnarray}}
\def\nn{\nonumber}

\def\tr{{\rm tr}\,}

\def\K{\mathcal{K}}

\newdimen\linethick  \linethick=0.4pt
\newdimen\hboxitspace    \hboxitspace=5pt
\newdimen\vboxitspace    \vboxitspace=5pt

\usepackage{calligra}
\DeclareMathAlphabet{\mathcalligra}{T1}{calligra}{m}{n}

\hoffset= - 1.3in         

\title{{\bf Vassiliev invariants for pretzel knots  } \vspace{.2cm}}

\author{ {\bf A. Sleptsov}\footnote{sleptsov@itep.ru}}
\date{ }

\begin{document}
 \maketitle

\vspace{-4.5cm}

\begin{center}
 \hfill ITEP/TH-36/15\\
 \hfill IITP/TH-21/15\\
\end{center}

\vspace{2.2cm}

\begin{center}
{\small {\it ITEP, Moscow 117218, Russia}}\\
{\small {\it Institute for Information Transmission Problems, Moscow 127994, Russia}}\\
{\small {\it National Research Nuclear University MEPhI, Moscow 115409, Russia }}\\
{\small {\it Laboratory of Quantum Topology,
Chelyabinsk State University, Chelyabinsk 454001, Russia}}
\end{center}

\vspace{1cm}

\centerline{ABSTRACT}

\bigskip

{\footnotesize We compute Vassiliev invariants up to order six for arbitrary pretzel knots, which depend on $g+1$ parameters $n_1,\ldots,n_{g+1}$. These invariants are symmetric polynomials in $n_1,\ldots,n_{g+1}$ whose degree coincide with their order. We also discuss their topological and integer-valued properties.}

\section{Introduction}
A set of Vassiliev invariants is conjectured to be a complete invariant of a knot as well as a set of colored quantum invariants. Despite these two sets were discovered approximately simultaneously (see \cite{vasset} and \cite{qinvset}), a progress in the construction and the calculation of colored quantum invariants is significantly greater than with Vassilev invariants. About calculations of quantum invariants during last years see, for example, \cite{qinvcalculus} and see \cite{ChD, DBSS1} for latest reviews about Vassiliev invariants. One of the most successful approach to quantum invariant calculations is to divide all knots into families and try to find explicit answers for them. It turns out in many cases that it is rather easy to compute or sometimes guess quantum invariants for particular family and the answer turns out to be amazingly simple and well-structured. This phenomenon can be illustrated by a famous family of torus knots whose all colored HOMFLY polynomials are given by the beautiful Rosso-Jones formula \cite{RJ}. This formula inspires many mathematicians to begin their research with torus knots. In particular, there were calculated Vassiliev invariants of torus knots up to order 6 in the paper \cite{Laba}. Recently \cite{pretzel1,pretzel2} it was obtained some explicit results for quantum invariants of pretzel knots which are a natural generalisation of simplest torus knots of a form $T[2,2n+1]$ to a Riemann surface of arbitrary genus $g$. It stimulates us to compute and discuss their Vassiliev invariants.

\section{Vassiliev invariants from Chern-Simons theory}

The incorporation of Vassiliev invariants in the path-integral representation is clear from the following picture.
Let $A$ be a connection on $\mathbb{R}^3$ taking values in some
representation $R$ of a Lie algebra $g$, i.e., in components:
$$
A=A_{i}^{a}(x)\,T^{a}\,dx^{i},
$$
where $T^{a}$ are the generators of $g$.
Let curve $C$ in $\mathbb{R}^3$ give a particular realization of knot $K$.
Consider the holonomy of $A$ along $C$, it is given by
the ordered exponent:
$$
\Gamma(C,A) = P \exp \oint\limits_{C} A = 1+ \oint\limits_{C} A_{i}^{a}(x) T^{a}+
\oint\limits_{C} A_{i_{1}}^{a_{1}}(x_{1})\int\limits_{0}^{x_{1}} A_{i_{2}}^{a_{2}}(x_{2})
\, T^{a_{1}} T^{a_{2}}+...
$$
The Wilson loop along $C$ is a function depending on $C$ and $A$ defined as a trace of holonomy:
$$ W_R(C,A)=\tr_R \Gamma(C,A) $$
According to \cite{WCS} there exists a functional $S_{CS} (A)$
(we write it down explicitly later) such that the integral averaging of the Wilson loop with the
weight $\exp\Big(-\frac{2\pi i}{\hbar} S(A)\Big)$  has the following remarkable property:
\be
\label{W}
\begin{array}{|c|}
\hline\\
\ \ \  \langle \, W_R(K)\,\rangle= \dfrac{1}{Z} \displaystyle \int DA \, \exp\Big(-\frac{2\pi i}{\hbar} S_{CS}(A)\Big)\, W_R(C,A) \ \ \  \\
\\
\hline
\end{array}
\ee
where
$$
 Z=\int DA\,\exp\Big(-\frac{2\pi i}{\hbar} S_{CS}(A) \Big)
$$
i.e. the averaging of $W_{R}(C,A)$ with the weight $\exp\Big(-\frac{2\pi i}{\hbar} S(A)\Big)$
does not depend on the realization $C$ of the knot in $\mathbb{R}^3$
but only on the topological class of equivalence of knot $K$ (in what follows we will denote
the averaging of quantity $Q$ with this weight by $\langle Q\rangle$) and therefore,  $\langle\, W(K)\,\rangle$ defines a knot invariant.

The distinguished Chern-Simons action giving the invariant average (\ref{W}) has the following form:
\be
\label{CSA}
S_{CS}(A)=\int \limits_{\mathbb{R}^3}\,\tr ( A\wedge dA +\dfrac{2}{3} A\wedge A\wedge A )
\ee
If we normalize the algebra generators $T^{a}$ as $\tr (T^{a}T^{b})=\delta^{a b}$ and define the structure constants $f$ of algebra $g$ as $[T^{a},T^{b}]=f_{a b c}\,T^{c}$  then the action takes the form:
$$
S_{CS}(A)=\epsilon^{i j k} \int\limits_{\mathbb{R}^3} dx^{3}\, A^{a}_{i} \partial_{j} A_{k}^{a} +\dfrac{1}{6} f_{a b c} A^{a}_{i}A^{b}_{j}A^{c}_{k}
$$

\begin{figure}[!ht]
\centering\leavevmode
\includegraphics[scale=0.75]{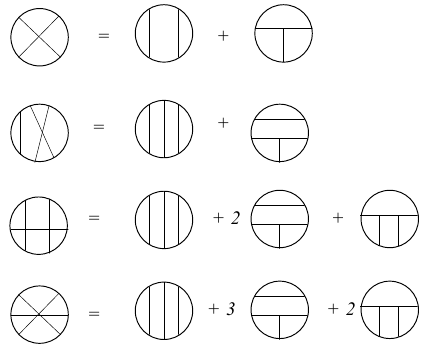}
\caption{Relation between trivalent diagrams and chord diagrams up to order 3}
\label{p_triv}
\end{figure}

\begin{figure}[h]
\centering\leavevmode
\includegraphics[width=10 cm]{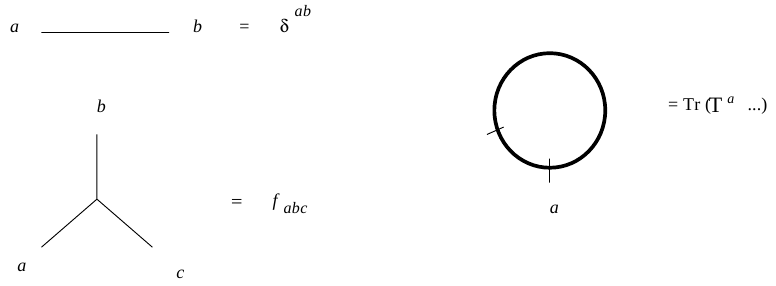}
\caption{Group-theoretical rules}
\label{rules}
\end{figure}

Formula (\ref{W}) is precisely the path integral representation of knot invariants. It is believed that all invariants of knots can be derived from this expression. Let us outline the appearance of Vassiliev invariants in this scheme. Obviously the mean value $\langle\,W(K)\,\rangle$ has the following structure:
\be
\label{eq1}
\nonumber
 \langle\, W(C,A)\, \rangle = \langle \,\sum\limits_{n=0}^{\infty} \oint dx_{1}\int
dx_{2}...\int dx_{n} A^{a_1}(x_{1})A^{a_2}(x_{2})...A^{a_3}(x_{n})\,
\tr(T^{a_1} T^{a_2}...T^{a_n}) \, \rangle=\\
=\sum\limits_{n=0}^{\infty} \oint dx_{1}\int
dx_{2}...\int dx_{n} \langle\,A^{a_1}(x_{1})A^{a_2}(x_{2})...A^{a_3}(x_{n})\,\rangle\,
\tr(T^{a_1} T^{a_2}...T^{a_n}) =
\sum\limits_{n=0}^{\infty} \sum \limits_{m=1}^{N_{n}} V_{n,m}\, G_{n,m}
 \ee
From this expansion we see that the information about the knot and the gauge
group enter in $\langle W(K) \rangle$ separately. The information about the embedding of a knot into
$\mathbb{R}^3$ is encoded in the integrals of the form:
$$
V_{n,m} \sim \oint dx_{1}\int
dx_{2}...\int dx_{n} \langle\,A^{a_1}(x_{1})A^{a_2}(x_{2})...A^{a_3}(x_{n})\,\rangle
$$
and the information about the gauge group and representation enter in the answer as the "group factors":
$$
G_{n,m} \sim \tr(T^{a_1} T^{a_2}...T^{a_n})
$$
$G_{k,m}$ are the group factors called \textit{chord diagrams with $n$ chords}. Chord diagrams with $n$ chords form a vector space $H_n$.
Despite $\langle W(K) \rangle$ being a knot invariant, the numbers $V_{n,m}$ are not invariants. This is because
the group elements $G_{n,m}$ are not independent, and the coefficients $V_{n,m}$ are invariants only up to relations among $G_{n,m}$. Dimensions of $H_{n}$ are summarized in the table:
\begin{equation}
\begin{array}{|c|c|c|c|c|c|c|}
\hline
n&1&2&3&4&5&6\\
\hline
dim(H_{n})&1&1&1&3&4&9\\
\hline
\end{array}
\end{equation}
In order to pass to Vassiliev invariants we have to choose some basis in the space of chord diagrams. We do it following \cite{La5}, refer to that paper for details. The so-called trivalent diagrams are introduced in a way represented for orders two and three in Figure \ref{p_triv}. Group-theoretical rules for graphical representation of chords and trivalent diagrams are presented in Figure \ref{rules}. For the general definition of trivalent diagrams refer to \cite{La5}, see also \cite{La6}.  

Let us explain the definition of trivalent diagrams on the first relation from Figure \ref{p_triv}: $$T^aT^bT^cT^d\delta^{ac}\delta^{bd}=T^aT^bT^aT^b=\mathpic{0.4}{\draw[black] (0,0) circle (1); \draw[-,black,thick] (-0.707106,0.707106) -- (0.707106,-0.707106); \draw[-,black,thick] (0.707106,0.707106) -- (-0.707106,-0.707106);}$$ 
$T^aT^bT^cT^d\delta^{ad}\delta^{bc}=T^aT^bT^bT^a=\mathpic{0.4}{\draw[black] (0,0) circle (1); \draw[-,black,thick] (-0.5,0.866025) -- (-0.5,-0.866025); \draw[-,black,thick] (0.5,0.866025) -- (0.5,-0.866025);}=T^aT^bT^aT^b-T^aT^bT^aT^b+T^aT^bT^bT^a=T^aT^bT^aT^b-T^aT^b\left(T^aT^b-T^bT^a\right)=T^aT^bT^aT^b-T^aT^b\left[T^aT^b\right]= T^aT^bT^aT^b-f^{abc}T^aT^bT^c=\mathpic{0.4}{\draw[black] (0,0) circle (1); \draw[-,black,thick] (-0.707106,0.707106) -- (0.707106,-0.707106); \draw[-,black,thick] (0.707106,0.707106) -- (-0.707106,-0.707106);} - \mathpic{0.4}{\draw[black] (0,0) circle (1); \draw[-,black,thick] (-1,0) -- (1,0); \draw[-,black,thick] (0,0) -- (0,-1);}$

\begin{figure}[h!]
\centering\leavevmode
\includegraphics[width=13 cm]{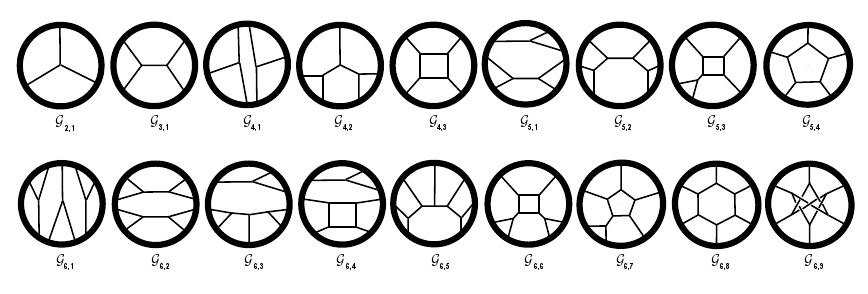}
\caption{Trivalent diagrams}
\label{p_chords}
\end{figure}

In Figure \ref{p_chords} one can find a collection of trivalent diagrams that form the so-called \textit{canonical basis} $\{{\cal{G}}_{ij} \}$ of $H_{n}$ up to order six. In the fundamental representation $R[1]$ their explicit expressions are given in the following table:
\begin{large}
\be
\label{trivd1}
\begin{array}{cc}
{\cal{G}}_{2,1} = -{1\over4}N^2+{1\over4} & {\cal{G}}_{6,1} = -{1\over64}N^6+{3\over64}N^4-{3\over64}N^2+{1\over64} \\ \\
{\cal{G}}_{3,1} = -{1\over8}N^3+{1\over8}N & {\cal{G}}_{6,2} = {1\over64}N^6-{1\over32}N^4+{1\over64}N^2 \\ \\
{\cal{G}}_{4,1} = {1\over16}N^4-{1\over8}N^2+{1\over16} & {\cal{G}}_{6,3} = {1\over64}N^6-{1\over32}N^4+{1\over64}N^2 \\ \\
{\cal{G}}_{4,2} = -{1\over16}N^4+{1\over16}N^2 & {\cal{G}}_{6,4} = -{1\over64}N^6+{3\over64}N^2-{1\over32} \\ \\
{\cal{G}}_{4,3} = {1\over16}N^4+{1\over16}N^2-{1\over8} & {\cal{G}}_{6,5} = -{1\over64}N^6+{1\over64}N^4 \\ \\
{\cal{G}}_{5,1} = {1\over32}N^5-{1\over16}N^3+{1\over32}N & {\cal{G}}_{6,6} = {1\over64}N^6+{1\over64}N^4-{1\over32}N^2 \\ \\
{\cal{G}}_{5,2} = -{1\over32}N^5+{1\over32}N^3 & {\cal{G}}_{6,7} = {1\over64}N^6-{1\over64}N^2 \\ \\
{\cal{G}}_{5,3} = {1\over32}N^5+{1\over32}N^3-{1\over16}N & {\cal{G}}_{6,8} = {1\over64}N^6+{1\over64}N^2-{1\over32} \\ \\
{\cal{G}}_{5,4} = {1\over32}N^5-{1\over32}N & {\cal{G}}_{6,9} = {3\over64}N^4-{5\over64}N^2+{1\over32} 
\end{array}
\ee
\end{large}

In the first symmetric representation $R=[2]$ they equal to
\begin{large}
\be
\label{trivd2}
\begin{array}{cc}
{\cal{G}}_{2,1} = -{1\over2}N^2-{1\over2}N+1 & {\cal{G}}_{6,1} = \left( {\cal{G}}_{2,1} \right)^3 \\ \\
{\cal{G}}_{3,1} = -{1\over4}N^3-{1\over4}N^2+{1\over2}N & {\cal{G}}_{6,2} = \left( {\cal{G}}_{3,1} \right)^2 \\ \\
{\cal{G}}_{4,1} = \left( {\cal{G}}_{2,1} \right)^2 & {\cal{G}}_{6,3} = {\cal{G}}_{2,1}\cdot {\cal{G}}_{4,2} \\ \\
{\cal{G}}_{4,2} = -{1\over8}N^4-{1\over8}N^3+{1\over4}N^2 & {\cal{G}}_{6,4} =  {\cal{G}}_{2,1}\cdot {\cal{G}}_{4,3}\\ \\
{\cal{G}}_{4,3} = {1\over8}N^4+{3\over8}N^3+N^2-{1\over8}N-2 & {\cal{G}}_{6,5} = -{1\over32}N^6-{1\over32}N^5+{1\over16} \\ \\
{\cal{G}}_{5,1} = {\cal{G}}_{2,1}\cdot {\cal{G}}_{3,1} & {\cal{G}}_{6,6} = {1\over32}N^6+{3\over32}N^5+{1\over4}N^4+{1\over8}N^3-{1\over2}N^2 \\ \\
{\cal{G}}_{5,2} = -{1\over16}N^5-{1\over16}N^4+{1\over8}N^3 & {\cal{G}}_{6,7} = {1\over32}N^6+{1\over8}N^5+{11\over32}N^4+{1\over8}N^3-{5\over8}N^2 \\ \\
{\cal{G}}_{5,3} = {1\over16}N^5+{3\over16}N^4+{1\over2}N^3+{1\over4}N^2+N & {\cal{G}}_{6,8} = {1\over32}N^6+{5\over32}N^5+{19\over32}N^4+{31\over32}N^3+{5\over8}N^2-{3\over8}N-2 \\ \\
{\cal{G}}_{5,4} = -{1\over16}N^5-{1\over4}N^4-{11\over16}N^3-{1\over4}N^2+{5\over4}N & {\cal{G}}_{6,9} = {7\over32}N^4+{9\over32}N^3-{11\over8}N^2-{9\over8}N+2 
\end{array}
\ee
\end{large}

Using this basis we rewrite (\ref{eq1}) through invariants:
\be
\label{B}
<W_{R}(K)>\,=\sum\limits_{n=0}^{\infty}\,\hbar^n\,\sum_{m=1}^{\dim(H_{n})}\,{\cal{V}}_{n,m}\,{\cal{G}}_{n,m}
\ee
Here ${\cal{V}}_{ij}$ are the so called finite-type or Vassiliev invariants of knots. They depend only on the knot under consideration but not on the group and its representation.

Now let us introduce primitive Vassiliev invariants.
It is a well known fact that the expansion of logarithm of any correlator in any QFT contains only connected Feynman diagrams (for more details about this situation in the Chern-Simons perturbation theory see \cite{La8}). This fact immediately leads to the following summation of
\be 
\label{C}
<W_{R}(K)>\,=\prod\limits_{n=0}^{\infty}\,\prod_{m=1}^{{\cal{N}}_n}\,
\exp\left(  \hbar^n{\cal{V}}^c_{n,m}\,{\cal{G}}^c_{n,m} \right),
\ee
where ${\cal{G}}^c$ are connected diagrams, ${\cal{V}}^c$ are primitive Vassiliev invariants. The Vassiliev invariants form a graded ring freely generated by primitive invariants. Here ${\cal{N}}_n$ is dimension of the space of connected chord diagrams (or equivalently the space of primitive Vassiliev invariants). The dimensions of these spaces up to order 6 are given in the following table:
\begin{equation}
\begin{array}{|c|c|c|c|c|c|c|}
\hline
n&1&2&3&4&5&6\\
\hline
{\cal{N}}_n&1&1&1&2&3&5\\
\hline
\end{array}
\end{equation}
The meaning of the expression (\ref{C}) is that ${\cal{V}}_{i,j}$ in (\ref{B}) are not independent. In fact only those coefficients ${\cal{V}}_{ij}$ are independent, for which the corresponding diagram ${\cal{G}}_{ij}$ is connected. Comparing $\hbar$ expansion of (\ref{C}) with (\ref{B}) we, for example, find:
\be
\label{vasrel0}
{\cal{V}}_{4,1}=&{1\over2}\, {\cal{V}}_{2,1}^2 \nn \\
{\cal{V}}_{5,1}=&{\cal{V}}_{2,1}{\cal{V}}_{3,1}, \nn \\
{\cal{V}}_{6,1}=&{1\over 6}{\cal{V}}_{2,1}^3, \nn\\
{\cal{V}}_{6,2}=&{1\over2}\, {\cal{V}}_{3,1}^2, \\
{\cal{V}}_{6,3}=&{\cal{V}}_{2,1}{\cal{V}}_{4,2}, \nn\\
{\cal{V}}_{6,4}=&{\cal{V}}_{2,1}{\cal{V}}_{4,3}.\nn 
\ee

The last but not least, from formulas (\ref{trivd2}) we can see that to compute Vassiliev invariants up to order 6 it is enough to have HOMFLY polynomials in the first symmetric representation $R=[2]$. And in the next section we provide corresponding formulas of HOMFLY polynomials for torus and pretzel knots.

\section{HOMFLY polynomials}

\subsection{Torus knots}
In the case of torus knots HOMFLY polynomials in all representations were calculated by Rosso and Jones in \cite{RJ}. So, let us consider torus knot $T[m,n]$ with mutually prime $m$ and $n$ and let $\chi_R$ are the Schur polynomials. We define the coefficients $c_R^Q$ from the relation
\be
\chi_R\{p^{(m)}\}\ = \sum_{Q\vdash m|R|} c_R^Q \chi_Q\{p\}
\label{cfromshift}
\ee
where
\be
p^{(m)}_k = p_{mk}
\label{pmk}
\ee
Thus, for the torus knot $T[m,n]$ one has
\be
H_R^{T[m,n]}\{p\} =
\sum_{Q\vdash m|R|} q^{-2\frac{n}{m}\varkappa_Q} c^Q_R \chi_Q^*, 
\label{Wrep}
\ee
where
\be
\chi_Q^* = \chi_Q\{ p_k = \frac{a^k-a^{-k}}{q^k-q^{-k}} \}.
\ee

This nice formula allows to compute HOMFLY polynomial in any representation.

\subsection{Pretzel knots}
These are knots and links formed by wrapping around a surface of genus $g$ without self-intersections, which can be different from $g=1$. The simplest set of this type has a knot diagram (see Figure 2), consisting of $g+1$ two-strand braids, and thus has $g+1$ different parameters $n_1,\ldots, n_{g+1}$ (for $g=1$ everything depends on the sum $n=n_1+n_2$). In literature (see \cite{pretz}) this family is known as the pretzel knots and links. The family is actually split into subfamilies, differing by mutual orientation of strands in the braids. Since we are interested in pretzel knots only, let us consider all possible configurations of parameters $n_1,\ldots, n_{g+1}$ and orientations, which provide only knots. There are 3 possible orientations: {\it antiparallel}, {\it parallel} and {\it mixed}.

\bigskip

{\bf Antiparallel} In the first case we put genus $g$ to be odd, all orientations of constituent braids must be antiparallel like on the picture
\begin{tikzpicture}[scale=0.5]
\draw[line width=1,white]
(1,1) -- (1,1);
\draw[line width=1]
(0,0) -- (1,0) -- (2,0.8) -- (3,0.8);
\draw[line width=1]
(0,0.8) -- (1,0.8);
\draw[line width=1]
(1,0.8) -- (1.375,0.5) (1.624,0.3)  -- (2,0) -- (3,0);
\draw[line width=1,->,-stealth] (0,0) -- (0.75,0);
\draw[line width=1,->,-stealth] (1,0.8) -- (0.25,0.8);
\end{tikzpicture}
all parameters $\overline {n_1},\dots,\overline {n_g}$ must be odd\footnote{Since for all qualities standing for the antiparallel case we use "bar", we denote parameters in this case as $\overline {n_1},\dots,\overline {n_{g+1}}$.}. Then we obtain a class of knots which we refer to {\it antiparallel} pretzel knots. Their HOMFLY polynomials in any symmetric representation are given as follows
\be
\label{homfa}
H^{\overline {n_1},\ldots,\overline  {n_{g+1}}}_R \ \  = \ \  \sum_{k=0}^r \Delta_k \, \prod_{i=1}^{g+1} \, \sum_{m=0}^r \bar{\bar a}_{km} \bar \lambda_m^{\bar n_i},
\ee
where $\Delta_k$ is {\it a quantum dimension} of the corresponding representation, $\bar \lambda_m$ is {\it an eigenvalue} of the corresponding R-matrix and $\bar{\bar a}_{km}$ is the corresponding {\it Racah coefficient}. Their values were computed in \cite{pretzel1},\cite{pretzel2} and can be listed as follows:
\be
\bar \lambda_m &=& \left(-q^{m-1}A\right)^m  \nn \\
\Delta_m &=& \chi^*_{[r+m,r-m]}={[2m+1]\over [r+m+1]![r-m]!}\prod_{i=0}^{2r-1}D_j\prod_{j=0}^{r-m-1}{D_{j-1}\over D_{r+m+j}}  \nn \\
\bar \Delta_m &=& D_{2m-1}\cdot \left(\prod_{j=0}^{m-2}\frac{D_j}{[j+2]} \right)^{\!\!2}\!\cdot D_{-1} \nn \\
{\bar{\bar a}}_{km} &=& {\bar\Delta_m\over\Delta_k}a_{mk} \nn \\
a_{km} &=& \alpha_{km} \, \cdot \, {\cal G} \nn \\
\alpha_{km} &=& (-1)^{r+k+m} [2m+1]\cdot {\Big([k]![m]!\Big)^2\,[r-k]!\,[r-m] !\over [r+k+1]!\,[r+m+1]!}
 \times \nn \\ &\times&
\ \sum_{j=\text{max}(r+m,r+k)}^{\text{min}(r+k+m,2r)} {(-1)^j\,[j+1]!\over [2r-j]!\ \Big([j-r-k]!\,[j-r-m]!\,[r+k+m-j]!\Big)^2} \nn \\
{\cal G} &=& \frac{G(r-m)\,G(j+1)}{G(r+k+1)\,G(j-r-m)} \nn \\
G(n) &=& \dfrac{1}{[n]!}\prod_{i=-1}^{n-2}D_{i}={(A/q;q)_n\over (q;q)_n}
\ee
and also we used here standard notations for quantum numbers $\{x\}= x-x^{-1},\ [x]=\frac{\{q^x\}}{\{q\}}$ and quantum factorials, for differentials $D_i = \frac{\{Aq^i\}}{\{q\}}$ and for symmetric q-Pochhammer symbols $(A;q)_n=\prod_{j=0}^{n-1}\{Aq^j\}$. Also note that at $A=q^N$, $G(n)$ becomes the q-binomial
$\left(\begin{array}{c} N+n-2\\n\end{array}\right)_q$.

\bigskip

{\bf Parallel} In the second case we put genus $g$ to be odd, all orientations of constituent braids must be parallel like on the picture
\begin{tikzpicture}[scale=0.5]
\draw[line width=1,white]
(1,1) -- (1,1);
\draw[line width=1]
(0,0) -- (1,0) -- (2,0.8) -- (3,0.8);
\draw[line width=1]
(0,0.8) -- (1,0.8);
\draw[line width=1]
(1,0.8) -- (1.375,0.5) (1.624,0.3)  -- (2,0) -- (3,0);
\draw[line width=1,->,-stealth] (0,0) -- (0.75,0);
\draw[line width=1,->,-stealth] (0,0.8) -- (0.75,0.8);
\end{tikzpicture}
all parameters $n_2,\ldots, n_{g+1}$ must be odd and $n_1$ must be even\footnote{We choose $n_1$ to be even for simplicity. In general, it is possible to choose any.}. Then we obtain a class of knots which we refer to {\it parallel} pretzel knots. Their HOMFLY polynomials in any symmetric representation are given as follows
\be
H^{n_1,\ldots,n_{g+1}}_{[r]}
 \ = \ \sum_{k=0}^r \, \bar \Delta_k \cdot \left\{ \ \prod_{i=1}^{g+1} \ \left( \sum_{m=0}^r a_{km} \, \lambda_{m}^{n_i} \right)  \right\}
\label{homfp}
\ee
where all constituents have similar meaning as in the previous case, explicit expression for $\lambda_{m}$ is the following
\be
\lambda_m = (-)^{m+1}{q^{-r^2+m^2+m}\over A^r}
\ee

\bigskip

{\bf Mixed} In the third case we put genus $g$ to be even, all orientations of constituent braids, except one, must be parallel, their corresponding parameters $n_2,\ldots, n_{g+1}$ must be odd, $\bar n_1$ must be even (again for simplicity distinguish $n_1$). Then we obtain a class of knots which we refer to {\it mixed} pretzel knots. Their HOMFLY polynomials in any symmetric representation are given as follows
\be
H^{\bar n_1, n_2,\ldots,n_{g+1}}_{[r]}
 \ = \ \sum_{k=0}^r \, \bar \Delta_k \cdot \left\{ \left(\sum_{m=0}^r \bar a_{km} \, \bar \lambda_{m}^{\bar n_1}\right) \ \prod_{i=2}^{g+1} \ \left( \sum_{m=0}^r a_{km} \, \lambda_{m}^{n_i} \right)  \ \right\} \ \
\label{homfm}
\ee

Thus, our formulas (\ref{homfa}), (\ref{homfp}) and (\ref{homfm}) provide the explicit answer for arbitrary pretzel knots in arbitrary symmetric representation. These formulas we can use to evaluate Vassiliev invariants in the next section.

\section{Vassiliev invariants}
In this section we first present the Vassilev invariants up to order 6 for torus knots evaluated by M. Alvarez and J. M. F. Labastida in \cite{Laba}, then we present our results for pretzel knots.

\subsection{Torus knots}
\be
v_{2,1}=& {1 \over 24}  \,( {n^2}-1 ) \,({m^2}-1),  \cr
v_{3,1}=& {1 \over 144} \,n\,m\,(  {n^2}-1 ) \,(  {m^2}-1 ), \cr
v_{4,2}=& {1 \over 240} \,( {n^2}-1 ) \, (m^2-1) \,
     ( 9\,{n^2}\,{m^2}-n^2-m^2-1 ), \cr
v_{4,3}=& {1 \over 240} \,(  {n^4}-1 ) \,(  {m^4}-1 ), \cr
\ee
\be
v_{5,2}=& {1 \over 28800} \,n\,m \,( n^2-1 )\, ( m^2-1) \,
     ( 69\,{n^2}\,{m^2}  - 21\,\big({n^2} + {m^2})-11\big), \cr
v_{5,3}=& {1 \over 57600} \,n\,m \,( n^2-1 )\, ( m^2-1) \,
     ( 11\,{n^2}\,{m^2} + {n^2} + {m^2}-9), \cr
v_{5,4}=& {1 \over 7200} \,n\,m\,(  {n^4}-1 ) \,( {m^4}-1 ), \cr
\ee
\be
v_{6,5}=& {1 \over 2520} \, (n^2-1) \, (m^2-1) \,
\big(  516\,{n^4}\,{m^4}- 289\,({n^2}\,{m^4}+{n^4}\,{m^2}) \cr &
\,\,\,\,\,\,\,\,\,\,\,\,\,\,\,\,\,\,\,\,\,\,\,\,\,\,\,\,\,\,\,\,
\,\,\,\,\,\,\,\,\,\,\,\,\,\,\,\,\,\,\,\,\,\,\,\,
- 44\,{n^2}\,{m^2} + 5\,({n^4}+m^4) + 5\,({n^2}+m^2) + 5 \big), \cr
v_{6,6}=& {1 \over 12096} \, (n^2-1) \, (m^2-1) \,
\big(53\,{n^4}\,{m^4}- 101\,({n^2}\,{m^4}+{n^4}\,{m^2}) \cr &
\,\,\,\,\,\,\,\,\,\,\,\,\,\,\,\,\,\,\,\,\,\,\,\,\,\,\,\,\,\,\,\,
\,\,\,\,\,\,\,\,\,\,\,\,\,\,\,\,\,\,\,\,\,\,\,\,
- 115\,{n^2}\,{m^2}- 24\,({n^4}+m^4) - 24\,({n^2}+m^2)  -24  \big), \cr
v_{6,7}=& {1 \over 10080} \, (n^2-1) \, (m^2-1) \,
\big( 419\,{n^4}\,{m^4}+ 209\,({n^2}\,{m^4}+{n^4}\,{m^2}) \cr &
\,\,\,\,\,\,\,\,\,\,\,\,\,\,\,\,\,\,\,\,\,\,\,\,\,\,\,\,\,\,\,\,
\,\,\,\,\,\,\,\,\,\,\,\,\,\,\,\,\,\,\,\,\,\,\,\,
- {n^2}\,{m^2} + 20\,({n^4}+m^4)  + 20\,({n^2}+m^2) +20 \big), \cr
v_{6,8}=& {1 \over 25200} \, (n^2-1) \, (m^2-1) \,
\big( 13\,{n^4}\,{m^4} + 13\,({n^2}\,{m^4}+{n^4}\,{m^2}) \cr &
\,\,\,\,\,\,\,\,\,\,\,\,\,\,\,\,\,\,\,\,\,\,\,\,\,\,\,\,\,\,\,\,
\,\,\,\,\,\,\,\,\,\,\,\,\,\,\,\,\,\,\,\,\,\,\,\,
+ 13\,{n^2}\,{m^2}  - 50\,({n^4}+m^4) - 50\,({n^2}+m^2) -50 \big), \cr
v_{6,9}=& {1 \over 5040} \, (n^2-1) \, (m^2-1) \,
\big( 31\,{n^4}\,{m^4} + 31\,({n^2}\,{m^4} + {n^4}\,{m^2}) \cr &
\,\,\,\,\,\,\,\,\,\,\,\,\,\,\,\,\,\,\,\,\,\,\,\,\,\,\,\,\,\,\,\,
\,\,\,\,\,\,\,\,\,\,\,\,\,\,\,\,\,\,\,\,\,\,\,\,
+ 31\,{n^2}\,{m^2} + 10\,({n^4}+m^4) + 10\,({n^2}+m^2) + 10 \big).
\cr
\ee

\subsection{Pretzel knots}
HOMFLY polynomials, obtained in the previous section, are symmetric under permutations of $\{n_i\}$. The reason is the following. Permutation of the two adjacent $n_i$'s is just a knot mutation. Since the HOMFLY polynomials in symmetric representations do not distinguish the mutant knots \cite{mut}, with the help of mutation one can permute $n_i \leftrightarrow n_{i+1}$. Vassiliev invariants up to order 6 also do not distinguish the mutant knots, hence their formulas have to include this symmetry. Taking this into account together with they are polynomials in $\{n_i\}$, we conclude that Vassiliev invariants can be expressed in terms of symmetric polynomials. In other words, we can choose some basis in the space of symmetric polynomials and express Vassiliev invariants in terms of basis elements with some coefficients  depending on genus $g$. Schur polynomials provide a distinguished basis in the space of symmetric polynomials, so we use it for our computations.

Below we present our results for three subfamilies of pretzel knots. To avoid notation ambiguities we use different letters standing for Vassiliev invariants for different subfamilies.

\subsubsection{Antiparallel}
\be
v_{2,1}  &=& \chi_{[1,1]} +{\text{g} \over 2}     \\
v_{3,1}  &=& -{1 \over 2}\left( \chi_{[2,1]} + 2\chi_{[1,1,1]} + \text{g}\chi_{[1]} \right)   
\\
v_{4,2}  &=& {1 \over 6}\left( \chi_{[3,1]} + 5\chi_{[2,2]} + 8\chi_{[2,1,1]} + 5\chi_{[1,1,1,1]} + {3 \over 2}\text{g}\chi_{[2]} + \left(9\text{g}+4\right) {1 \over 2}\chi_{[1,1]} +{\text{g} \over 4}\left(3\text{g}+2\right) \right)  
\\
v_{4,3}  &=& {1 \over 12}\left( 3\chi_{[2,2]} + 3\chi_{[2,1,1]} - 3\chi_{[1,1,1,1]} + 4\chi_{[1,1]} +{\text{g} \over 2} \right)  
\\
v_{5,2}  &=& -{1 \over 24}\left( \chi_{[4,1]} + 21\chi_{[3,2]} + 25\chi_{[3,1,1]} + 63\chi_{[2,2,1]} + 57\chi_{[2,1,1,1]} + 14\chi_{[1,1,1,1,1]} +  \right. \nn \\ &+& \left.  2\text{g}\chi_{[3]} + 4\left(2+7\text{g}\right)\chi_{[2,1]} + 2\left(12+13\text{g}\right)\chi_{[1,1,1]} + \text{g}\left(6\text{g}+5\right)\chi_{[1]} \right) 
\\
v_{5,3}  &=& -{1 \over 6}\left( \chi_{[3,2]} + \chi_{[3,1,1]} + 3\chi_{[2,2,1]} + 3\chi_{[2,1,1,1]} - 2\chi_{[1,1,1,1,1]} + {3 \over 2}\text{g}\chi_{[2,1]} + 4\chi_{[1,1,1]} + \text{g}\chi_{[1]} \right)    
\\
v_{5,4}  &=& -{1 \over 6}\left( \chi_{[3,2]} + \chi_{[3,1,1]} + 3\chi_{[2,2,1]} - 2\chi_{[1,1,1,1,1]} + 2\chi_{[2,1]} + 2\chi_{[1,1,1]} \right)  
\\
v_{6,5}  &=& {1 \over 360}\left( 3\chi_{[5,1]} + 192\chi_{[4,2]} + 405\chi_{[3,3]} + 207\chi_{[4,1,1]} + 1578\chi_{[3,2,1]} + 1113\chi_{[3,1,1,1]} + 1050\chi_{[2,2,2]} + 2097\chi_{[2,2,1,1]} + \right. \nn \\ &+& \left.  1332\chi_{[2,1,1,1,1]} + 81\chi_{[1,1,1,1,1,1]} + {15 \over 2}\text{g}\chi_{[4]} + {15 \over 2}\left(47\text{g}+8\right)\chi_{[3,1]} + 90\left(7\text{g}+2\right)\chi_{[2,2]} + {15 \over 2}\left(145\text{g}+64\right)\chi_{[2,1,1]} + \right. \nn \\ &+& \left.  {15 \over 2}\left(71\text{g}+104\right)\chi_{[1,1,1,1]} + {15 \over 2}\text{g}\left(15\text{g}+8\right)\chi_{[2]} + {3 \over 2}\left(135\text{g}^2+170\text{g}-32\right)\chi_{[1,1]} + {3 \over 2}\text{g}\left(10\text{g}^2+15\text{g}+6\right) \right)  
\\
v_{6,6}  &=& {1 \over 360}\left( - 35\chi_{[3,3]} - 70\chi_{[3,2,1]} + 85\chi_{[3,1,1,1]} - 50\chi_{[2,2,2]} + 225\chi_{[2,2,1,1]} + 470\chi_{[2,1,1,1,1]} - 80\chi_{[1,1,1,1,1,1]} + \right. \nn \\ &+& \left.  20\left(3\text{g}-4\right)\chi_{[3,1]} + 20\left(6\text{g}-11\right)\chi_{[2,2]} + 10\left(21\text{g}-31\right)\chi_{[2,1,1]} + 10\left(3\text{g}+41\right)\chi_{[1,1,1,1]} + \right. \nn \\ &+& \left.  {15 \over 2}\text{g}\left(3\text{g}+1\right)\chi_{[2]} + {3 \over 2}\left(15\text{g}^2+85\text{g}-98\right)\chi_{[1,1]} + 3\text{g}\left(5\text{g}+3\right) \right)
\\
v_{6,7}  &=& {1 \over 360}\left( 75\chi_{[4,2]} + 255\chi_{[3,3]} + 75\chi_{[4,1,1]} + 810\chi_{[3,2,1]} + 375\chi_{[3,1,1,1]} + 525\chi_{[2,2,2]} + 585\chi_{[2,2,1,1]} - 210\chi_{[2,1,1,1,1]} - \right. \nn \\ &-& \left.  330\chi_{[1,1,1,1,1,1]} + 60\left(\text{g}+2\right)\chi_{[3,1]} + 60\left(2\text{g}+7\right)\chi_{[2,2]} + 60\left(2\text{g}+13\right)\chi_{[2,1,1]} - 60\left(\text{g}-5\right)\chi_{[1,1,1,1]} + \right. \nn \\ &+& \left.  45\text{g}\chi_{[2]} + \left(105\text{g}+124\right)\chi_{[1,1]} + 2\text{g}  \right)
\\
v_{6,8}  &=& {1 \over 144}\left( 3\chi_{[4,2]} - 5\chi_{[3,3]} + 3\chi_{[4,1,1]} + 2\chi_{[3,2,1]} - 5\chi_{[3,1,1,1]} + 7\chi_{[2,2,2]} + 15\chi_{[2,2,1,1]} - 4\chi_{[2,1,1,1,1]} - 20\chi_{[1,1,1,1,1,1]} - \right. \nn \\ &-& \left.  8\chi_{[2,2]} + 8\chi_{[3,1]} + 4\chi_{[2,1,1]} + 28\chi_{[1,1,1,1]} - 10\chi_{[1,1]} - \text{g}  \right) 
\\
v_{6,9}  &=& {1 \over 720}\left( 15\chi_{[4,2]} + 95\chi_{[3,3]} + 15\chi_{[4,1,1]} + 250\chi_{[3,2,1]} + 95\chi_{[3,1,1,1]} + 155\chi_{[2,2,2]} + 75\chi_{[2,2,1,1]} - 140\chi_{[2,1,1,1,1]} + \right. \nn \\ &+& \left.  20\chi_{[1,1,1,1,1,1]} + 40\chi_{[3,1]} + 200\chi_{[2,2]}  + 260\chi_{[2,1,1]}  - 100\chi_{[1,1,1,1]} + 86\chi_{[1,1]} + 3\text{g} \right)    
\ee

Let us note that any $\chi_{\Delta}$ depends on genus $g$, because it depends on $g+1$ variables $\{n_1,\ldots,n_{g+1}\}$. However some coefficients of $\chi_{\Delta}$ in the formulas above do not depend on $g$. Actually, we can make the following three observations:
\begin{enumerate}
\item coefficients of leading terms do not depend on $g$, i.e. they are constants;
\item coefficients in $v_{5,4}$ are constants;
\item three following combinations have constant coefficients:
\be
120\,v_{6,9} - v_{2,1} \nn \\
12\,v_{4,3} - v_{2,1}  \\
72\,v_{6,8} + v_{2,1} \nn 
\ee
\end{enumerate}

\subsubsection{Parallel}
\be
u_{2,1}  &=& \left( n_1^2 + 2\chi_{[1]}n_1 + \chi_{[2]}-\chi_{[1,1]} - \text{g} \right) {1 \over 2}  \\
u_{3,1}  &=& -\left( n_1^3 + 3\chi_{[1]}n_1^2 + n_1\left(3\chi_{[2]} - \left(3\text{g}-1\right) {1 \over 2} \right) + \left(\chi_{[3]}-\chi_{[2,1]}+\chi_{[1,1,1]} - \chi_{[1]}\right)  \right) {1 \over 3}  
\\
u_{4,2}  &=& \left( {14 \over 3}\left(\chi_{[4]}-\chi_{[3,1]}+\chi_{[2,1,1]}-\chi_{[1,1,1,1]}\right) - {16 \over 3}\left(\chi_{[2]}-\chi_{[1,1]}\right) + {2\text{g} \over 3} \, + \right. \nn \\ &+& \left.  
n_1\cdot {8 \over 3}\left(7\chi_{[3]}-\chi_{[2,1]}+\chi_{[1,1,1]} - \left(3\text{g}+1\right)\chi_{[1]}\right) \ \ + \ \ n_1^2\cdot{4 \over 3}\left(21\chi_{[2]}+9\chi_{[1,1]} -2\left(3\text{g}-1\right)\right) \, + \right. \nn \\ &+& \left.  
n_1^3\cdot{56 \over 3}\chi_{[1]}n_1^3 \ \  + \ \ {14 \over 3}n_1^4 \right) {1 \over 16}  
\\
u_{4,3}  &=& \left( {2 \over 3}\left(\chi_{[4]}-\chi_{[3,1]}+\chi_{[2,1,1]}-\chi_{[1,1,1,1]}\right) -  {2\text{g} \over 3} + {8 \over 3}\left(\chi_{[3]}-\chi_{[2,1]}+\chi_{[1,1,1]}\right)n_1 \, + \right. \nn \\ &+& \left. 
n_1^2\cdot4\left(\chi_{[2]}+\chi_{[1,1]}\right) \ \  + \ \  n_1^3\cdot{8 \over 3}\chi_{[1]} \ \  + \ \ n_1^4\cdot{2 \over 3} \right) {1 \over 16}  
\ee
\be
u_{5,2}  &=& \left( -51\left( \chi_{[5]} - \chi_{[4,1]} + \chi_{[3,1,1]} - \chi_{[2,1,1,1]} + \chi_{[1,1,1,1,1]} \right) + 70\left( \chi_{[3]} - \chi_{[2,1]} + \chi_{[1,1,1]} \right) - 19\chi_{[1]} \, + \right. \nn \\ &+& \left.  
n_1\cdot\left(-255\chi_{[4]}+45\chi_{[3,1]}+30\chi_{[2,2]}-45\chi_{[2,1,1]}+15\chi_{[1^4]} + 30\left(4{+}3\text{g}\right)\chi_{[2]}-30\chi_{[1,1]}-\left({45 \over 2}\text{g}^2{-}15\text{g}{+}{23 \over 2}\right)  \right)+\right. \nn \\ &+& \left.  n_1^2\left(-510\chi_{[3]}-255\chi_{[2,1]}-60\chi_{[1,1,1]}  +  15\left(-1+15\text{g}\right)\chi_{[1]} \right)+n_1^3\left(-510\chi_{[2]}-300\chi_{[1,1]} + 35\left(-1+3\text{g}\right) \right) +\right. \nn \\ &+& \left.    n_1^4\left( -255\chi_{[1]} \right) - 51n_1^5   \right){1 \over 180}  
\\
u_{5,3}  &=& \left( -9\left( \chi_{[5]} - \chi_{[4,1]} + \chi_{[3,1,1]} - \chi_{[2,1,1,1]} + \chi_{[1,1,1,1,1]} \right) + 10\left( \chi_{[3]} - \chi_{[2,1]} + \chi_{[1,1,1]} \right) - \chi_{[1]} +\right. \nn \\ &+& \left.  
n_1\cdot\left(-45\chi_{[4]}+15\chi_{[3,1]}+30\chi_{[2,2]}-15\chi_{[2,1,1]}-15\chi_{[1,1,1,1]}+30\chi_{[2]}+30\chi_{[1,1]} + \left(-16+15\text{g}\right) \right)+\right. \nn \\ &+& \left.  
n_1^2\cdot\left(-90\chi_{[3]}-45\chi_{[2,1]} + 15\left(-1+3\text{g}\right)\chi_{[1]} \right) \ \ + \ \ n_1^3\cdot\left(-90\chi_{[2]}-60\chi_{[1,1]} + 5\left(-1+3\text{g}\right)\right) + \right. \nn \\ &+& \left.   n_1^4\cdot\left( -45\chi_{[1]} \right) \ \ - \ \  9\cdot n_1^5 \right) {1 \over 180}  
\\
u_{5,4}  &=& -\left(  \chi_{[5]} - \chi_{[4,1]} + \chi_{[3,1,1]} - \chi_{[2,1^3]} + \chi_{[1^5]}  - \chi_{[1]} + n_1\cdot\left( 5\left( \chi_{[4]} - \chi_{[3,1]} + \chi_{[2,1,1]} - \chi_{[1^4]} \right) + 10\chi_{[1,1]} - 1 \right) +\right. \nn \\ &+& \left.  
n_1^2\cdot\left( 10\chi_{[3]} + 5\chi_{[2,1]} + 10\chi_{[1,1,1]} \right) \ \ + \ \ n_1^3\cdot\left( 10\chi_{[2]} + 10\chi_{[1,1]} \right) \ \ + \ \ n_1^4\cdot\left( 5\chi_{[1]} \right) \ \ + \ \ n_1^5 \right) {1 \over 30}  
\ee
\be
u_{6,5}  &=& \left( 203\left( \chi_{[6]}{-}\chi_{[5,1]}{+}\chi_{[4,1,1]}{-}\chi_{[3,1^3]}{+}\chi_{[2,1^4]}{-}\chi_{[1^6]} \right)  -340\left( \chi_{[4]}{-}\chi_{[3,1]}{+}\chi_{[2,1,1]}{-}\chi_{[1^4]} \right) + 140\left( \chi_{[2]}{-}\chi_{[1,1]} \right) - 3\text{g} \, +  \right. \nn \\ &+& \left.  
n_1 \cdot \left(1218\chi_{[5]}-198\chi_{[4,1]}-180\chi_{[3,2]}+198\chi_{[3,1,1]}+60\chi_{[2,2,1]}-78\chi_{[2,1^3]}+ 18\chi_{[1^3]} - 20\left(47+21\text{g}\right)\chi_{[3]} \, +\right.\right. \nn \\  &&  \left.\left. \hspace{5.2cm} + \,  20\left(11+3\text{g}\right)\chi_{[2,1]}-20\left(5+3\text{g}\right)\chi_{[1,1,1]}+10\left(9\text{g}^2+6\text{g}+13\right)\chi_{[1]}\right) + \right. \nn \\ &+& \left.   
n_1^2 \cdot \left(3045\chi_{[4]}+1635\chi_{[3,1]}+360\chi_{[2,2]}+345\chi_{[2,1,1]}+15\chi_{[1,1,1,1]}- 510\left(1+3\text{g}\right)\chi_{[2]}- 30\left(-1+21\text{g}\right)\chi_{[1,1]}+\right.\right. \nn \\ && \left.\left. \hspace{12.25cm} + 5\left(45\text{g}^2-33\text{g}+16\right)\right) + \right. \nn \\ &+& \left.  n_1^3 \cdot \left(4060\chi_{[3]} +3440\chi_{[2,1]}+820\chi_{[1,1,1]}-20\left(-13+81\text{g}\right)\chi_{[1]}\right) +\right. \nn \\ &+& \left.   n_1^4 \cdot \left(3045\chi_{[2]}+2025\chi_{[1,1]} -170\left(-1+3\text{g}\right)\right)  \ \ + \ \ 1218\chi_{[1]}\cdot n_1^5 \ \ + \ \ 203\cdot n_1^6 \right) {1 \over 720}  
\\ 
u_{6,6}  &=& \left( 5\left( \chi_{[6]}{-}\chi_{[5,1]}{+}\chi_{[4,1,1]}{-}\chi_{[3,1^3]}{+}\chi_{[2,1^4]}{-}\chi_{[1^6]}\right) -30\left( \chi_{[4]}-\chi_{[3,1]}+\chi_{[2,1,1]}-\chi_{[1^4]} \right)+19\left( \chi_{[2]}-\chi_{[1,1]} \right) + 6\text{g}  +  \right. \nn \\ &+& \left.  n_1\cdot\left(30\chi_{[5]}+60\chi_{[4,1]}-60\chi_{[3,1,1]}-30\chi_{[3,2]}-30\chi_{[2,2,1]}+120\chi_{[2,1,1,1]}-90\chi_{[1,1,1,1,1]}- 30\left(3+\text{g}\right)\chi_{[3]}-\right.\right. \nn \\ && \left.\left. \hspace{6.44cm} - 30\left(6-\text{g}\right)\chi_{[2,1]}-30\left(-1+\text{g}\right)\chi_{[1,1,1]}+2\left(4+15\text{g}\right)\chi_{[1]}\right)  + \right. \nn \\ &+& \left. 
n_1^2\cdot\left(75\chi_{[4]}+120\chi_{[3,1]}+30\chi_{[2,2]}- 165\chi_{[2,1,1]} + \left(45\text{g}^2-75\text{g}+68\right) \frac12 -    45\left(1+3\text{g}\right)\chi_{[2]}-45\left(5+\text{g}\right)\chi_{[1,1]}\right)  +\right. \nn \\ &+& \left.  
n_1^3\cdot\left(100\chi_{[3]}+5\chi_{[2,1]} -200\chi_{[1,1,1]} - 30\left(-1+5\text{g}\right)\chi_{[1]}\right)  +\right. \nn \\ &+& \left.  
n_1^4\cdot\left(75\chi_{[2]}-15\chi_{[1,1]}-15\left(-1+3\text{g}\right)\right) \ \ + \  \ 30\chi_{[1]}\cdot n_1^5 \ \ + \ \ 5\cdot n_1^6  \right) {1 \over 360}  
\\
u_{6,7}  &=& \left( 18\left( \chi_{[6]}{-}\chi_{[5,1]}{+}\chi_{[4,1,1]}{-}\chi_{[3,1^3]}{+}\chi_{[2,1^4]}{-}\chi_{[1^6]} \right) -10\left( \chi_{[4]}-\chi_{[3,1]}+\chi_{[2,1,1]}-\chi_{[1^4]} \right)-7\chi_{[2]} + 7\chi_{[1,1]}-\text{g}  +  \right. \nn \\ &+& \left.  n_1\cdot\left(108\chi_{[5]}-78\chi_{[4,1]}+78\chi_{[3,1,1]}-30\chi_{[3,2]}+30\chi_{[2,2,1]}-78\chi_{[2,1,1,1]}+ 48\chi_{[1,1,1,1,1]}- 40\chi_{[3]}+ \right.\right. \nn \\ && \left.\left. \hspace{7.45cm} +  130\chi_{[2,1]}-40\chi_{[1,1,1]}-2\left(-8+15\text{g}\right)\chi_{[1]}\right)  +  \right. \nn \\ &+& \left.  n_1^2\cdot\left(270\chi_{[4]}+75\chi_{[3,1]}-30\chi_{[2,2]}+150\chi_{[2,1,1]}+ 15\chi_{[1^4]} -  15\left(1+3\text{g}\right)\chi_{[2]} -45\left(\text{g}-3\right)\chi_{[1,1]}-\left(15\text{g}-1\right) {1 \over 2} \right)  +  \right. \nn \\ &+& \left.  
n_1^3\cdot\left(360\chi_{[3]}+375\chi_{[2,1]} + 240\chi_{[1,1,1]}-20\left(-1+3\text{g}\right)\chi_{[1]}\right)  +  \right. \nn \\ &+& \left.  n_1^4\cdot\left(270\chi_{[2]}+240\chi_{[1,1]}-5\left(-1+3\text{g}\right) \right)  \ \ + \ \  108\chi_{[1]}\cdot n_1^5 \ \ + \ \  18\cdot n_1^6 \right) {1 \over 180}  
\ee
\be
u_{6,8}  &=& \left( 4\left( \chi_{[6]}-\chi_{[5,1]}+\chi_{[4,1,1]}-\chi_{[3,1,1,1]}+\chi_{[2,1,1,1,1]}-\chi_{[1,1,1,1,1,1]} \right) - 24\chi_{[2]}+24\chi_{[1,1]}+20\text{g}  + \right. \nn \\ &+& \left.   
n_1\cdot\left(24\left( \chi_{[5]}-\chi_{[4,1]}+\chi_{[3,1,1]}-\chi_{[2,1,1,1]}+\chi_{[1,1,1,1,1]} \right)+240\chi_{[2,1]}-48\chi_{[1]}\right) + \right. \nn \\ &+& \left. 
n_1^2\cdot\left(60\chi_{[4]}+60\chi_{[3,1]}+240\chi_{[2,2]}+ 300\chi_{[2,1,1]}-180\chi_{[1,1,1,1]}+240\chi_{[1,1]}-24\right)+ \right. \nn \\ &+& \left. 
n_1^3\cdot\left(80\chi_{[3]}+40\chi_{[2,1]}+80\chi_{[1,1,1]}\right) \ \ + \ \ 
n_1^4\cdot\left(60\chi_{[2]}+60\chi_{[1,1]}\right) \ \ + \ \ 24\chi_{[1]}\cdot n_1^5 \ \ + \ \ 4\cdot n_1^6  \right) {1 \over 2880} 
\\
u_{6,9}  &=& \left(  3\left( \chi_{[6]}-\chi_{[5,1]}+\chi_{[4,1,1]}-\chi_{[3,1,1,1]}+\chi_{[2,1,1,1,1]}-\chi_{[1,1,1,1,1,1]} \right)-2\chi_{[2]}+2\chi_{[1,1]}-\text{g} + \right. \nn \\ &+& \left. 
n_1\cdot\left(18\left( \chi_{[5]}-\chi_{[4,1]}+\chi_{[3,1,1]} - \chi_{[2,1,1,1]}+\chi_{[1,1,1,1,1]} \right)+20\chi_{[2,1]}-4\chi_{[1]}\right) + \right. \nn \\ &+& \left. n_1^2\cdot\left(45\chi_{[4]}+5\chi_{[3,1]}- 20\chi_{[2,2]}+ 25\chi_{[2,1,1]}+ 25\chi_{[1,1,1,1]}+20\chi_{[1,1]}-2\right)+\right. \nn \\ &+& \left.  n_1^3\cdot\left(60\chi_{[3]}+70\chi_{[2,1]}+60\chi_{[1,1,1]}\right) \ \ +  \ \
n_1^4\cdot\left(45\chi_{[2]}+45\chi_{[1,1]}\right) \ \ + \ \ 18\chi_{[1]}\cdot n_1^5 \ \ + \ \ 3\cdot n_1^6  \right) {1 \over 240}
\ee

In this case we also have the following:
\begin{enumerate}
\item coefficients of leading terms do not depend on $g$, i.e. they are constants;
\item coefficients in $u_{5,4}$ are constants;
\item three following combinations have constant coefficients:
\be
120\,u_{6,9} - u_{2,1} \nn \\
12\,u_{4,3} - u_{2,1}  \\
72\,u_{6,8} + u_{2,1} \nn 
\ee
\end{enumerate}

\subsubsection{Mixed}
\be
w_{2,1}  &=& \left(  \chi_{[2]}-\chi_{[1,1]} - \text{g} + n_1\left(  -2\chi_{[1]}  \right)   \right) {1 \over 2} 
\\
w_{3,1}  &=& \left( 2\left( \chi_{[3]}-\chi_{[2,1]}+\chi_{[1,1,1]} \right) - 2\chi_{[1]}  +  n_13\left( -2\chi_{[2]} + \text{g} \right)  +  n_1^23\chi_{[1]} \right) {1 \over 6} 
\\
w_{4,2}  &=& \left( 7\left( \chi_{[4]}-\chi_{[3,1]}+\chi_{[2,1,1]}-\chi_{[1,1,1,1]} \right) - 8\left( \chi_{[2]}-\chi_{[1,1]} \right) + \text{g}  + \right. \nn \\ &+& \left.   n_1\cdot\left( -28\chi_{[3]}+4\chi_{[2,1]}-4\chi_{[1,1,1]} + 4\left(4+3\text{g}\right)\chi_{[1]} \right) \ \  + \ \   n_1^2\cdot\left( 30\chi_{[2]}+18\chi_{[1,1]} - 6\text{g} \right) \ \ - \ \ n_1^3\cdot4\chi_{[1]} \right) {1 \over 24} 
\\
w_{4,3}  &=& \left(  \chi_{[4]}-\chi_{[3,1]}+\chi_{[2,1,1]}-\chi_{[1,1,1,1]}  - \text{g}  + \right. \nn \\ &+& \left.   n_1\cdot\left( -4\left(\chi_{[3]}-\chi_{[2,1]}+\chi_{[1,1,1]}\right) + 8\chi_{[1]} \right) \ \   + \ \ n_1^2\cdot\left( 6\chi_{[2]}+6\chi_{[1,1]} \right)   \right) {1 \over 24} 
\\
w_{5,2}  &=& \left( 102\left( \chi_{[5]}-\chi_{[4,1]}+\chi_{[3,1,1]}-\chi_{[2,1,1,1]}+\chi_{[1,1,1,1,1]} \right) - 140\left( \chi_{[3]}-\chi_{[2,1]}+\chi_{[1,1,1]} \right) + 38\chi_{[1]}  +  \right. \nn \\ &+& \left.    
n_1\cdot\left( -510\chi_{[4]}+90\chi_{[3,1]}+60\chi_{[2,2]}-90\chi_{[2,1,1]}+30\chi_{[1,1,1,1]} + 60\left(8+3\text{g}\right)\chi_{[2]} - 15\text{g}\left(3\text{g}+4\right)\, \right)  +  \right. \nn \\ &+& \left.    
n_1^2\cdot\left( 810\chi_{[3]}+540\chi_{[2,1]}+90\chi_{[1,1,1]} - 180\left(1+2\text{g}\right)\chi_{[1]} \right)  +  \right. \nn \\ &+& \left.   n_1^3\cdot\left(-360\chi_{[2]}-300\chi_{[1,1]} + 30\text{g} \right)  \ \ + \ \   n_1^4\cdot15\chi_{[1]} \right) {1 \over 360} 
\\
w_{5,3}  &=& \left( 9\left( \chi_{[5]}-\chi_{[4,1]}+\chi_{[3,1,1]}-\chi_{[2,1,1,1]}+\chi_{[1,1,1,1,1]} \right) - 10\left( \chi_{[3]}-\chi_{[2,1]}+\chi_{[1,1,1]} \right) + \chi_{[1]}  +  \right. \nn \\ &+& \left.    
n_1\cdot\left( -45\chi_{[4]}+15\chi_{[3,1]}+30\chi_{[2,2]}-15\chi_{[2,1,1]}-15\chi_{[1,1,1,1]} + 60\chi_{[2]} - 15\text{g} \right)  +  \right. \nn \\ &+& \left.    
n_1^2\cdot\left( 75\chi_{[3]}+60\chi_{[2,1]}-15\chi_{[1,1,1]} - 45\text{g}\chi_{[1]} \right) \ \ + \ \ n_1^3\cdot\left(-30\chi_{[2]}-30\chi_{[1,1]} \right)  \right) {1 \over 180} 
\\
w_{5,4}  &=& \left(  \chi_{[5]}-\chi_{[4,1]}+\chi_{[3,1,1]}-\chi_{[2,1,1,1]}+\chi_{[1,1,1,1,1]} -  \chi_{[1]}  +  \right. \nn \\ &+& \left.    
n_1\cdot\left( -5\left(\chi_{[4]}-\chi_{[3,1]}+\chi_{[2,1,1]}-\chi_{[1,1,1,1]}\right) + 10\chi_{[2]} \right)  +   \right. \nn \\ &+& \left.
n_1^2\cdot\left( 10\chi_{[3]}+5\chi_{[2,1]}+10\chi_{[1,1,1]} - 10\chi_{[1]} \right) \ \  + \ \    n_1^3\cdot\left(-5\chi_{[2]}-5\chi_{[1,1]} \right)  \right) {1 \over 30} 
\\
w_{6,5}  &=& \left( 203\left( \chi_{[6]}{-}\chi_{[5,1]}{+}\chi_{[4,1,1]}{-}\chi_{[3,1^3]}{+}\chi_{[2,1^4]}{-}\chi_{[1^6]} \right)-340\left( \chi_{[4]}{-}\chi_{[3,1]}{+}\chi_{[2,1,1]}{-}\chi_{[1^4]} \right)+ 140\left( \chi_{[2]}-\chi_{[1,1]} \right) - 3\text{g}  +  \right. \nn \\ &+& \left.  n_1\cdot\left(-1218\chi_{[5]}+198\chi_{[4,1]}+180\chi_{[3,2]}-198\chi_{[3,1,1]}-60\chi_{[2,2,1]}+78\chi_{[2,1,1,1]}-18\chi_{[1,1,1,1,1]} +  \right.\right. \nn \\ && \left.\left. \hspace{0.34cm}  + 20\left(74+21\text{g}\right)\chi_{[3]}-20\left(2+3\text{g}\right)\chi_{[2,1]}+20\left(2+3\text{g}\right)\chi_{[1,1,1]}-2\left(45\text{g}^2+ 180\text{g}+86\right)\chi_{[1]}\right)  +   \right. \nn \\ &+& \left.   n_1^2\cdot\left(2535\chi_{[4]}+1725\chi_{[3,1]}+420\chi_{[2,2]}+255\chi_{[2,1,1]}+45\chi_{[1^4]}- \right.\right. \nn \\ && \left.\left. \hspace{2mm} -\,  30\left(44+45\text{g}\right)\chi_{[2]}-30\left(20+21\text{g}\right)\chi_{[1,1]}+15\text{g}\left(12\text{g}+7\right)\right)  +  \right. \nn \\ &+& \left.    n_1^3\cdot\left(-1960\chi_{[3]}-2300\chi_{[2,1]}-640\chi_{[1,1,1]}+20\left(8+33\text{g}\right)\chi_{[1]}\right)  + \right. \nn \\ &+& \left. n_1^4\cdot\left(405\chi_{[2]}+375\chi_{[1,1]}-15\text{g}\right) \ \  - \ \  6\chi_{[1]}\cdot n_1^5 \right) {1 \over 720} 
\\
w_{6,6}  &=& \left( 5\left( \chi_{[6]}{-}\chi_{[5,1]}{+}\chi_{[4,1,1]}{-}\chi_{[3,1^3]}{+}\chi_{[2,1^4]}{-}\chi_{[1^6]} \right) -30\left( \chi_{[4]}{-}\chi_{[3,1]}{+}\chi_{[2,1,1]}{-}\chi_{[1^4]} \right) +  19\left( \chi_{[2]}-\chi_{[1,1]} \right) + 6\text{g} \, + \right. \nn \\ &+& \left.  n_1\cdot\left(-30\chi_{[5]}-60\chi_{[4,1]}+30\chi_{[3,2]}+60\chi_{[3,1,1]}+30\chi_{[2,2,1]}-120\chi_{[2,1,1,1]} +90\chi_{[1,1,1,1,1]} + 10\left(3\text{g}-1\right)\chi_{[3]} \,- \right.\right. \nn \\ && \left.\left. \hspace{6cm} - \,10\left(-4+3\text{g}\right)\chi_{[2,1]}+10\left(-13+3\text{g}\right)\chi_{[1,1,1]}- 6\left(-2+5\text{g} \right)\chi_{[1]}\right)  +  \right. \nn \\ &+& \left.    n_1^2\cdot\left(30\chi_{[4]}+135\chi_{[3,1]}+60\chi_{[2,2]}-180\chi_{[2,1,1]}-15\chi_{[1^4]}-15\left(9\text{g}-16\right)\chi_{[2]}-45\left(\text{g}-2\right)\chi_{[1,1]}+\frac{15\text{g}}{2}\left(3\text{g}-1\right) \right)  +  \right. \nn \\ &+& \left.    n_1^3\cdot\left(15\chi_{[3]}+30\chi_{[2,1]}+135\chi_{[1,1,1]}+20\left(-4+3\text{g}\right)\chi_{[1]}\right) \right) {1 \over 360} 
\\
w_{6,7}  &=& \left( \left( \chi_{[6]}{-}\chi_{[5,1]}{+}\chi_{[4,1,1]}{-}\chi_{[3,1^3]}{+}\chi_{[2,1^4]}{-}\chi_{[1^6]} \right) -20\left( \chi_{[4]}-\chi_{[3,1]}+\chi_{[2,1,1]}-\chi_{[1^4]} \right)-  14\left( \chi_{[2]}-\chi_{[1,1]} \right)-2\text{g}  \, +  \right. \nn \\ &+& \left. n_1\cdot\left(-216\chi_{[5]}+156\chi_{[4,1]}+60\chi_{[3,2]}-156\chi_{[3,1,1]}-60\chi_{[2,2,1]}+156\chi_{[2,1^3]}- 96\chi_{[1^5]}+  \right.\right. \nn \\ && \left.\left.  \hspace{4.92cm} +\,380\chi_{[3]}-80\chi_{[2,1]} +140\chi_{[1^3]} -12\left(4+5\text{g}\right)\chi_{[1]}\right)  +  \right. \nn \\ &+& \left. 
n_1^2\cdot\left(510\chi_{[4]}+180\chi_{[3,1]}-60\chi_{[2,2]}+ 270\chi_{[2,1,1]} +60\chi_{[1^4]}-30\left(22+3\text{g}\right)\chi_{[2]}-30\left(10+3\text{g}\right) \chi_{[1,1]} +45\text{g}\right)  +  \right. \nn \\ &+& \left. 
n_1^3\cdot\left(-450\chi_{[3]}-600\chi_{[2,1]}- 330\chi_{[1,1,1]}+ 60\left(2+\text{g}\right)\chi_{[1]}\right) \ \  + \ \  75n_1^4\cdot\left(\chi_{[2]}+\chi_{[1,1]}\right) \right) {1 \over 360}
\ee
\be
w_{6,8}  &=& \left( \chi_{[6]}-\chi_{[5,1]}+\chi_{[4,1,1]}-\chi_{[3,1,1,1]}+\chi_{[2,1,1,1,1]}-\chi_{[1,1,1,1,1,1]}-6\left( \chi_{[2]}+\chi_{[1,1]} \right)+5\text{g}  \,  +   \right. \nn \\ &+& \left.  
n_1\cdot\left(-6\chi_{[5]}+6\chi_{[4,1]}-6\chi_{[3,1,1]}+6\chi_{[2,1,1,1]}-6\chi_{[1,1,1,1,1]}+20\chi_{[3]}+40\chi_{[2,1]}+20\chi_{[1,1,1]}-4\chi_{[1]}\right)  +   \right. \nn \\ &+& \left.   
n_1^2\cdot\left(15\chi_{[4]}+15\chi_{[3,1]}+60\chi_{[2,2]}+75\chi_{[2,1,1]}-45\chi_{[1,1,1,1]}+60\chi_{[1,1]}\right)  +  \right. \nn \\ &+& \left. n_1^3\cdot\left(-10\chi_{[3]}+40\chi_{[2,1]}-10\chi_{[1,1,1]}+40\chi_{[1]}\right) \ \  + \ \   15n_1^4\cdot\left(\chi_{[2]}+\chi_{[1,1]}\right) \right) {1 \over 720} 
\\
w_{6,9}  &=& \left( 9\left( \chi_{[6]}-\chi_{[5,1]}+\chi_{[4,1,1]}-\chi_{[3,1,1,1]}+\chi_{[2,1,1,1,1]}-\chi_{[1,1,1,1,1,1]} \right)-6\left( \chi_{[2]}+\chi_{[1,1]} \right)-3\text{g} \, +  \right. \nn \\ &+& \left.  
n_1\cdot\left(-54\chi_{[5]}+54\chi_{[4,1]}-54\chi_{[3,1,1]}+54\chi_{[2,1,1,1]}-54\chi_{[1,1,1,1,1]}+100\chi_{[3]}-40\chi_{[2,1]}+100\chi_{[1,1,1]}-52\chi_{[1]}\right)+  \right. \nn \\ &+& \left.  
n_1^2\cdot\left(135\chi_{[4]}+15\chi_{[3,1]}-60\chi_{[2,2]}+75\chi_{[2,1,1]}+75\chi_{[1,1,1,1]}-240\chi_{[2]}-180\chi_{[1,1]}\right)+  \right. \nn \\ &+& \left.  n_1^3\cdot\left(-130\chi_{[3]}-200\chi_{[2,1]}-130\chi_{[1,1,1]}+40\chi_{[1]}\right) \ \ + \ \ 15n_1^4\cdot\left(\chi_{[2]}+\chi_{[1,1]}\right) \right) {1 \over 720}
\ee

In this case we also have the following:
\begin{enumerate}
\item coefficients of leading terms do not depend on $g$, i.e. they are constants;
\item coefficients in $w_{5,4}$ are constants;
\item three following combinations have constant coefficients:
\be
120\,w_{6,9} - w_{2,1} \nn \\
12\,w_{4,3} - w_{2,1}  \\
72\,w_{6,8} + w_{2,1} \nn 
\ee
\end{enumerate}
Thus, we see that these three observations are valid for all pretzel subfamilies, i.e. they are universal for any pretzel knot. It is very promising, probably, it helps to find a distinguished basis in the space of chord diagrams, because the current one (trivalent diagrams) is accidental.

\section{Properties of the Vassiliev invariants}
\subsection{Distinguishing knots}
How many of the Vassiliev invariants are needed to distinguish pretzel knots? In the case of torus knots the answer was found in \cite{Laba}. The Vassiliev invariants of the second and third orders are enough. In the case of pretzel knots the answer is unknown at the present moment. We definitely know that only $v_{2,1}^{\K}$ is not enough. For example, knots $(3,3,3)$ and $(-3,5,21)$ have same second Vassiliev invariants but different HOMFLY polynomials.

\subsection{Topological information}
Which Vassiliev invariants contain topological information? In other words we are looking for relations among them additional to (\ref{vasrel0}). In the case of torus knots there are only one independent Vassiliev invariant at each order up to order 6 \cite{Laba}. In the case of pretzel knots we found the only relation at order 6 only for antiparallel subfamily:
\be
-103v_{2, 1}+240v_{4, 3}+1080v_{6, 6}-180v_{6, 7}+630v_{6, 8}+4770v_{6, 9}-60v_{2, 1}^2 + \nn \\+90v_{2, 1}^3-180v_{3, 1}^2-180v_{2, 1}v_{4, 2}-1080v_{2, 1}v_{4, 3} \ =  0.
\ee
There are no more universal relations. We can say that antiparallel pretzel subfamily contains less topological information than two others. This feature is rather surprising and deserves futher studies.

\subsection{Integer-valued}
{\it These results are valid for all families of pretzel knots.}

Let us rescale Vassiliev invariants by normalization on the trefoil
\be
\tilde v^{\K}_{i,j} = \dfrac{v^{\K}_{i,j}}{v^{3_1}_{i,j}}
\ee
and multiply them on the following factors
\be
\tilde v_{2, 1}, \ \tilde v_{3, 1}, \   31\tilde v_{4, 2}, \ 5\tilde v_{4, 3}, \  11\tilde v_{5, 2}, \ \tilde v_{5, 3}, \ \tilde v_{5, 4}, \  5071\tilde v_{6, 5}, \ 29\tilde v_{6, 6}, \ 1531\tilde v_{6, 7}, \ 17\tilde v_{6, 8}, \ 271\tilde v_{6,9},
\ee
then such defined Vassiliev invariants take only integer values for all knots. For orders $i=2,3,4$ we can prove it by the straightforward enumerations, for $i=5,6$ we have a lot of numerical results.

\be
v^{3_1}_{2,1} = 4, \ v^{3_1}_{3,1} = -8, \ v^{3_1}_{4,2} = {62\over3}, \ v^{3_1}_{4,3} = {10\over3}, \  v^{3_1}_{5,2} = -{176\over3}, \ v^{3_1}_{5,3} = -{32\over3}, \ v^{3_1}_{5,4} = -8, \\ v^{3_1}_{6,5} = {5071\over30}, \ v^{3_1}_{6,6} = {58\over15}, \ v^{3_1}_{6,7} = {3062\over45}, \ v^{3_1}_{6,8} = {17\over18}, \ v^{3_1}_{6,9} = {271\over30}
\ee



\vspace{2cm}

\section*{Acknowledgements}

Our work is partly supported by grants NSh-1500.2014.2, by RFBR grants 13-02-00457, by the joint grants 15-52-50034-YaF, 15-51-52031-NSC-a, by 14-01-92691-Ind-a, 
14-02-31446-Mol-a and 15-31-20832-Mol-a-ved. Also we are partly supported by the Quantum Topology Lab of Chelyabinsk State University (Russian Federation government grant 14.Z50.31.0020).

\end{document}